\documentclass[11pt]{article}
\usepackage{graphicx,caption2,subfigure}
\makeatletter
\newcommand{\singlespacing}{\let\CS=\@currsize\renewcommand{\baselinestretch}{1.0}\tiny\CS}
\newcommand{\doublespacing}{\let\CS=\@currsize\renewcommand{\baselinestretch}{1.5}\tiny\CS}

\begin{document}
\title {Paradigm Shifts and a Possible Resolution of the PAMELA-Paradox in Astroparticle Physics.}
\author { Goutam
Sau$^1$\thanks{e-mail:sau$\_$goutam@yahoo.com}, S. K.
Biswas$^2$\thanks{e-mail: sunil$\_$biswas2004@yahoo.com}$\&$ S.
Bhattacharyya$^3$\thanks{e-mail: bsubrata@www.isical.ac.in
(Communicating Author).}\\
{\small $^1$ Beramara RamChandrapur High School,}\\
 {\small South 24-Pgs,743609(WB),India.}\\
{\small $^2$ West Kodalia Adarsha Siksha Sadan, New Barrackpore,}\\
 {\small Kolkata-700131, India.}\\
  {\small $^3$ Physics
and Applied Mathematics Unit(PAMU),}\\
 {\small Indian Statistical Institute, Kolkata - 700108, India.}}
\date{}
\maketitle
\bigskip
\doublespacing
\begin{abstract}
Very recently, PAMELA Collaboration has formally reported two sets
of data on positron and antiproton flux measurements done at very
high energies and with unprecedented accuracy. The reports reveal a
puzzle of great topical interest and importance : it is decisively
found that there is a sharp smooth rise of the positron fraction,
whereas for antiproton production no such rises occur; rather the
fraction either flattens, or shows signs of falling off gradually
with increasing energies. The present work is just an attempt to
decipher the riddle with the help of a host of radically new ideas
about the particle-structure, the multiparticle-production
mechanisms and the concept of nucleon break-downs into the
constituent partons. The application of these ideas found remarkable
successes in the past; exactly similar or more striking are the
findings by the present study.

\bigskip
 \par Keywords: Cosmic-ray interactions with the Earth, Positron emission, Cosmic-rays high energy interactions, Dark Matter. \\
\par PACS nos.: 94.94., 79.20.Mb., 13.85.Tp., 95.35.+d.
\end{abstract}
\newpage
\doublespacing
\section{Introduction and Background :}
Since the end of October 2008, there has been a tremendous surge of
interest and excitement, and then a consequent flurry of activity
among the astroparticle and high energy physicists. The two works
that shook the world of Astroparticle and Cosmic Ray physics in the
last one fortnight are the startling revelations made by Adriani et
al in two consecutive reports\cite{Adriani1,Adriani2} : (i) The
PAMELA satellite experiment by Adriani et al\cite{Adriani1} observed
and convincingly demonstrated a sharply rising ratio of the positron
flux measurements upto the studied 10-100GeV range of (secondary)
energy; (ii) conversely, the findings on the antiproton-to-proton
flux showed no such similar behaviour; rather the ratio-values
showed strong tendencies of flattening in the range of 80-100GeV
secondary energy\cite{Adriani2}. This striking contrast in the
nature of antiparticle-to-particle ratios poses a serious puzzle to
the theorists and occupies the centre-stage of astroparticle physics
domain today, stimulating more than two dozens of works within a
very very short span \cite{Hamed1}-\cite{Huh1}. In fact, some
previous studies\cite{Brun1,Picozza1} had already given very careful
but strong hints to the possibilities of such discoveries finally
reported only very recently, for which the paradox, once formally
reported, instantly caught such a large number of physicists to a
feverish pitch compelling them to act.
\par In the domain of astroparticle physics, particularly with regard
to the studies on dark matter (DM) and the weakly interacting
massive particles (WIMPs), searches for antimatter cosmic rays
comprising positrons and antiprotons constitute a very important
corner. Generically, the spectra of both positrons and antiprotons
are expected to fall with increasing energy, relative to the
corresponding matter particles which are here obviously electrons
and protons respectively. Any deviation from such standard
expectations might be a strong indication of any new primary cosmic
rays\cite{Serpico1,Cholis1}. In fact, such discoveries might unravel
new windows to the physics of dark matter (DM) and/or provide
valuable clues to the sources of ultrahigh energy cosmic rays, both
of which are still thoroughly unknown. These factors explain the
reasons for being drawn to such intense activities by the physicists
on the issue in question which really remains an enigma to the
adherents of the Standard Models (SMs) in High Energy Physics,
Astroparticle Physics and Cosmology, with all their ramifications
and interconnectedness.
\par In the present work here we will concentrate on understanding
the nature of the positron flux ratio alone in the light of some
non-standard ideas, hypotheses and ansatzs about the structure of
particles, the mechanism for particle interactions and finally the
mode of multiple production of hadrons. Very interestingly, the
other ratio of $\overline{P}/P$ turns, in the light of the model(s)
applied here, to a non-issue. Because, there is an exclusivity on
the production of positrons arising out of the putatively novel
concepts about the structure of hadrons. This helps us to obtain
with some non-standard additional and asymmetric sources for
$e^+$-production whereas, for the antiprotons no such asymmetric
source exists.
\par Before digressing, at the very outset, we would like to make a few comments.
 Most surprisingly, the results are no wonder to us; because we
had appreciated and
emphasised\cite{Bhattacharyya1}-\cite{Bhattacharyya3} the importance
and impact of such similar findings through a published work roughly
twenty years ago\cite{Bhattacharyya1} in a chain of related works;
based entirely on the same new paradigms which are outlined in the
next section in some detail. The present work is just the
resurrection of some of our past
works\cite{Bhattacharyya1}-\cite{Bhattacharyya3} with the latest and
newly obtained data from PAMELA collaboration. So neither we build
up any new model nor we refurbish the old model. We simply apply an
old model built up by one of the authers (SB) to explain the
characteristics of the new data produced by PAMELA collaboration on
excess production of positrons. By way, PAMELA is the acronym for
``Payload for Antimatter-Matter Exploration and Light-nuclei
Astrophysics"\cite{Barger1}.
\section{Paradigm Shifts : Brief Outlines}
\par(i) In the realm of Particle Physics we would introduce the
concept integer-charged partonic constituents for hadrons (both
mesons and baryons) with an old five-parton model. And the partons,
from the viewpoint of these radical
works\cite{Bhattacharyya4}-\cite{Bhattacharyya5} are identified to
be the muonic leptons like positively and negatively charged muons
and the muonic neutrinos. So, borrowing the phrase from Baek and
Ko\cite{Baek1}, the hadrostructures here are really `leptophilic'
or, more precisely, `muonophilic'. The details are to be obtained
from the works by one of us (SB)\cite{Bhattacharyya4} and the
references therein.
\par (ii) In the fields of Astroparticle and Cosmic Ray phenomena we
introduced the concept of nucleon-breaking mechanism, while the
ultra high energy cosmic rays pass through and collide with
intergalactic medium consisting of various light nuclei in highly
energised states. This concept is entirely different from the ideas
of proton- or neutron-decay. In fact, our previous studies on the
nature of positron spectrum and estimation of positron flux
fractions were based on these two sets of new ideas which mark a
radical departure from the Standard Model(SM)-based points of view.
\par(iii) Our approach to the studies on this particular problem of
excess production of positron fractions pertains in no way to the
existence or annihilation of cosmological dark
matter(DM)\cite{Steffen1} - cold or hot, with spin zero, unity or
else. Nor the little Higgs scalars are of any concern to us.
\par (iv) Besides, none of the Standard Model-related ideas like
weakly interacting massive particles (WIMPs), supersymmetric
particles (SPs) or supersymmetry, for that matter, and the
Kaluza-Klein particles in extra dimensions would be our concern
here.
\par (v) Regarding our choice on the propagation model for the ultra
high energy cosmic rays we will avoid the complicated ones and
accept the simplest and most commonly used model, called Simple
Leaky Box Model (SLBM).
\section{The Excess Positron Fraction : The Masterformula and The Results }
Based on the above-stated assumptions, ansatzs and arguments, we
deduced roughly two decades ago the formula for excess production of
positrons in one of our previous
works\cite{Bhattacharyya1}-\cite{Bhattacharyya3}. So, in order to
prevent the repetitive presentations of the same calculational
steps, we will simply extract\cite{Bhattacharyya1} for our present
purpose the final working expression therefrom given below.
\par  The standard sources
for electrons and positrons are given by
\par $\pi^-$ $\rightarrow$ $\mu^-$ $\rightarrow$ $e^-$
\par $\pi^+$ $\rightarrow$ $\mu^+$ $\rightarrow$ $e^+$
\par where the pions shown here are the secondaries abundantly
produced in nature by multiple production of hadrons known as
multiparticle phenomena. Similarly, secondary protons and
antiprotons are also produced by multiparticle mechanism. In fact,
this is the only method for production of the proton-antiprotons.
\par As stated above, the standard source of cosmic electron production is the normal
route of negatively charged pion-decays, where pions of all
varieties are produced by multiple production of hadron mechanism.
There is no other extraneous source of electron production. But for
production of cosmic ray positrons the additional, exotic and
non-standard source is the proposed nucleon-breaking mechanism with
positive muons residing inside the structure of protons as its
constituents. Through the proposed proton-breaking mechanism
supposed to be operational only at superhigh energies positive muons
are set free, which then could produce positrons through their
normal decay channel. These twin (standard and non-standard) sources
contribute to the positrons. The model- based final expression for
positron fraction, defined as the ratio of the total flux over total
electron plus positron flux, is given by\cite{Bhattacharyya1}
\begin{equation}
    \frac{\phi_{e^+}}{\phi_{e^+}+\phi_{e^-}}=\frac{1}{2+C'E^{0.5}_{e^+}\sin^2\theta_{cut}}+\beta E^{0.5}_{e^+}
\end{equation}
where $C'$ is a parameter related to the physics of proton-breaking
mechanism, $\theta_{cut}$ is the cut-off angle of emission or
detection of the positrons with the vertical. In the experiments by
Adriani et al the cut-off angle is not precisely mentioned for which
we have calculated for both small-angle $(0-10^{0})$(Fig.1) and
large-angle (Fig.2) emissions of the positrons. $\beta$ is a
parameter which is to be chosen by the methods of fitting the data.
We have maintained the same value $\beta$ for both small-angle and
large-angle scattering. The parameter values are given in Table-1
and Table-2.
\begin{figure}[h]
\subfigure[]{
\begin{minipage}{.5\textwidth}
\centering
\includegraphics[width=2.5in]{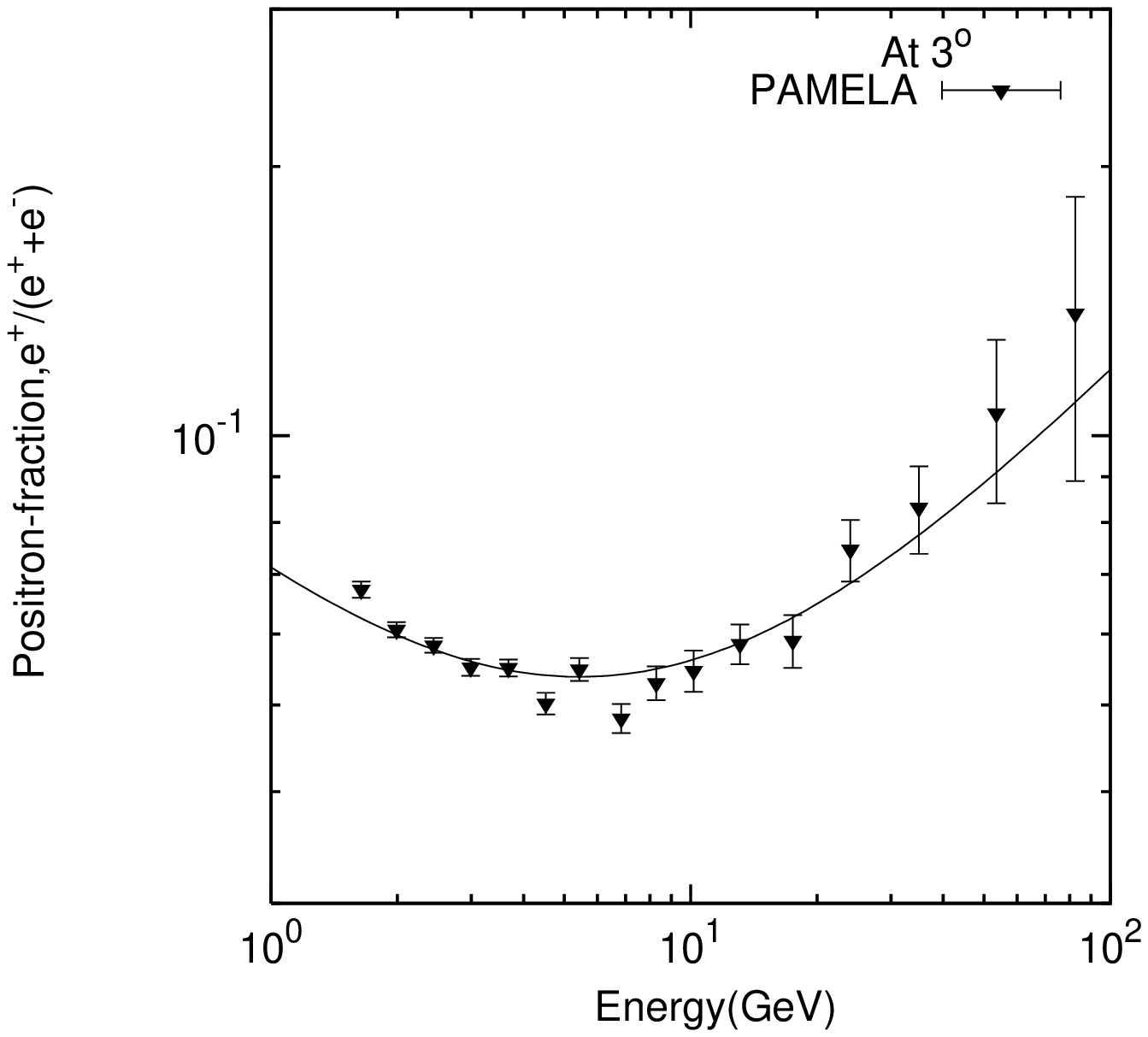}
\setcaptionwidth{2.6in}
\end{minipage}}%
\subfigure[]{
\begin{minipage}{0.5\textwidth}
\centering
 \includegraphics[width=2.5in]{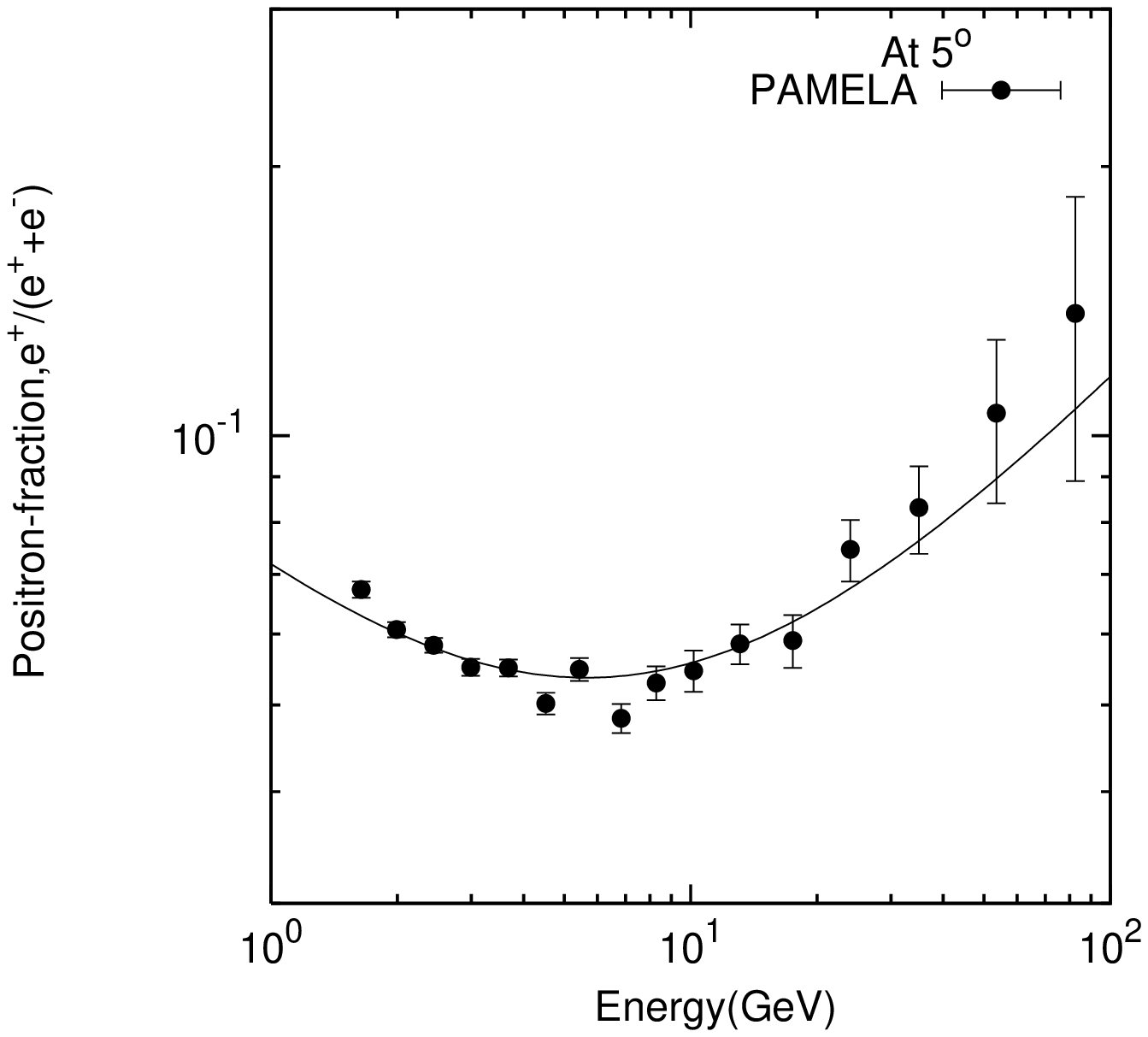}
 \end{minipage}}%
\vspace{0.01in} \subfigure[]{
\begin{minipage}{0.5\textwidth}
\centering
\includegraphics[width=2.5in]{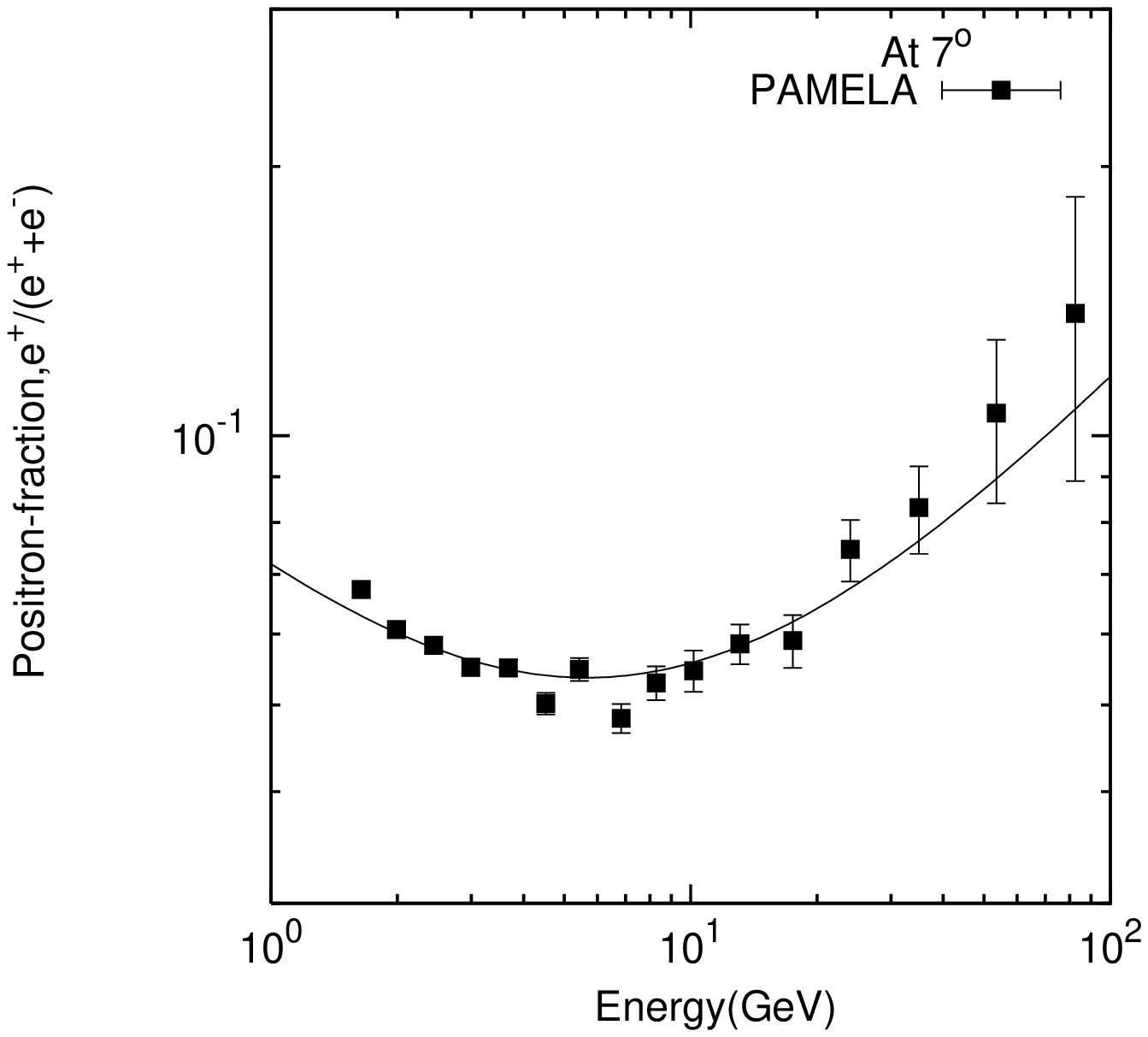}
\end{minipage}}%
\subfigure[]{
\begin{minipage}{.5\textwidth}
\centering
 \includegraphics[width=2.5in]{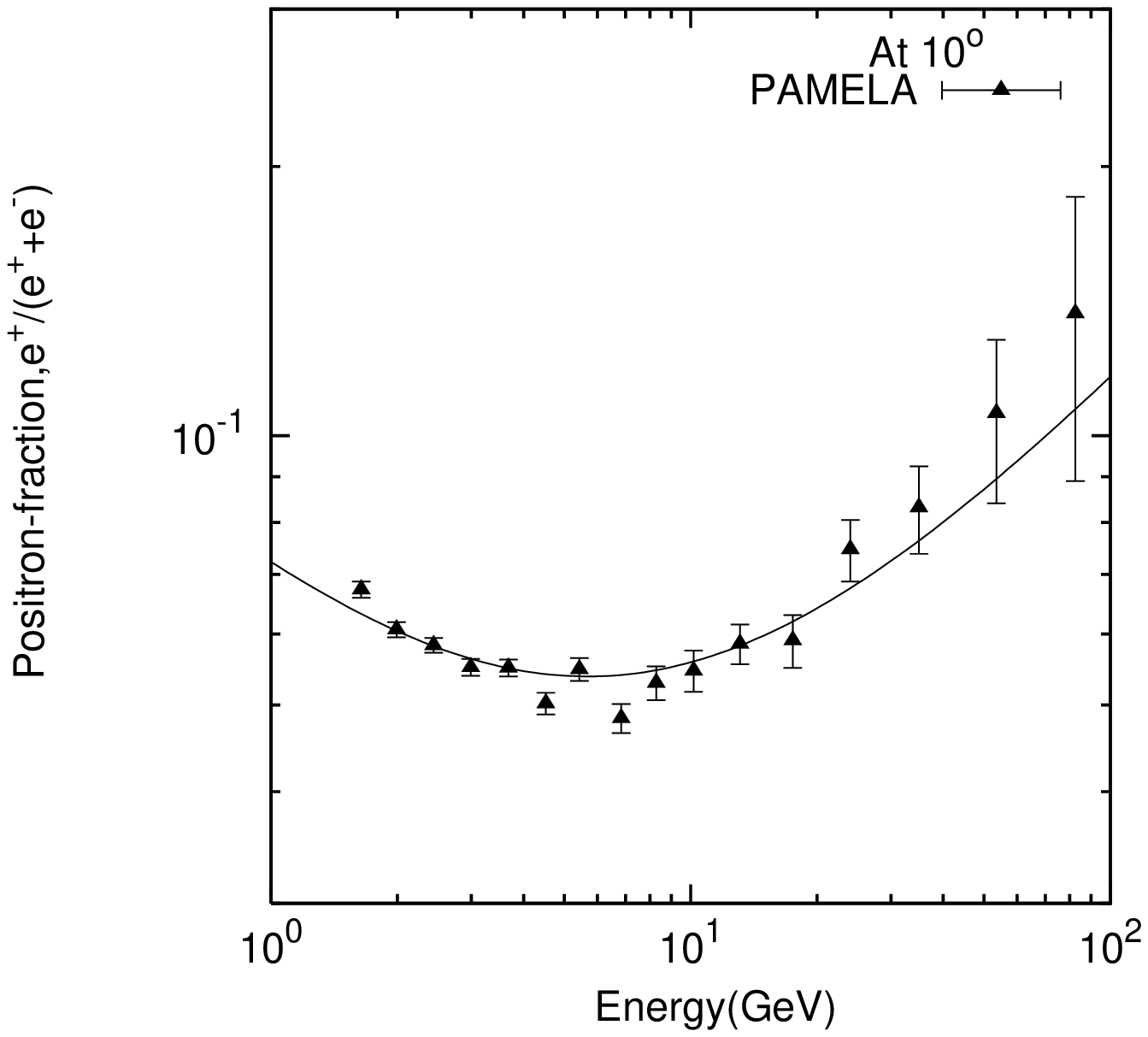}
 \end{minipage}}%
\caption{Data points are taken from Ref.\cite{Adriani1}. The errors are only statistical. The solid curve shows
the results based on the present working formula $(eqn.1)$.}
\end{figure}
\begin{table}[t]
\begin{center}
\begin{small}
\caption{Chosen numerical values of the fit parameters in the
expression for the positron fraction[with small-angle emission].}
\begin{tabular}{|c|c|c|c|}\hline
 $\theta_{cut}$ & $c'$ & $\beta$ & $\frac{\chi^2}{ndf}$\\
 \hline
 $3^{o}$ & $5417.46\pm 56.45$ &$0.011\pm0.0002$  & $4.775/11$\\
 \hline
  $5^{o}$ & $1904.63 \pm 18.13$ & $0.011\pm0.0002$  & $4.242/11$ \\
 \hline
  $7^{o}$ & $968.567 \pm 9.219$ & $0.011\pm0.0002$  & $4.242/11$ \\
 \hline
  $ 10^{o}$ & $472.714 \pm 4.592$ & $0.011\pm0.0002$  & $3.271/10$ \\
 \hline
\end{tabular}
\end{small}
\end{center}
\end{table}
\par The results based on the calculations are
presented only by graphical plots,[Fig.1 and Fig.2] alongwith the
two adjoining Tables [Table-1 and Table-2] which provide the
necessary parameter values. The results are controlled and dominated
by the second factor, power-law based term in the final expression.
In the data-analysis at ultrahigh energies the first term plays no
significant role. And so the results virtually grow independence of
the angle of emission.

\begin{figure}[h]
\subfigure[]{
\begin{minipage}{.5\textwidth}
\centering
\includegraphics[width=2.5in]{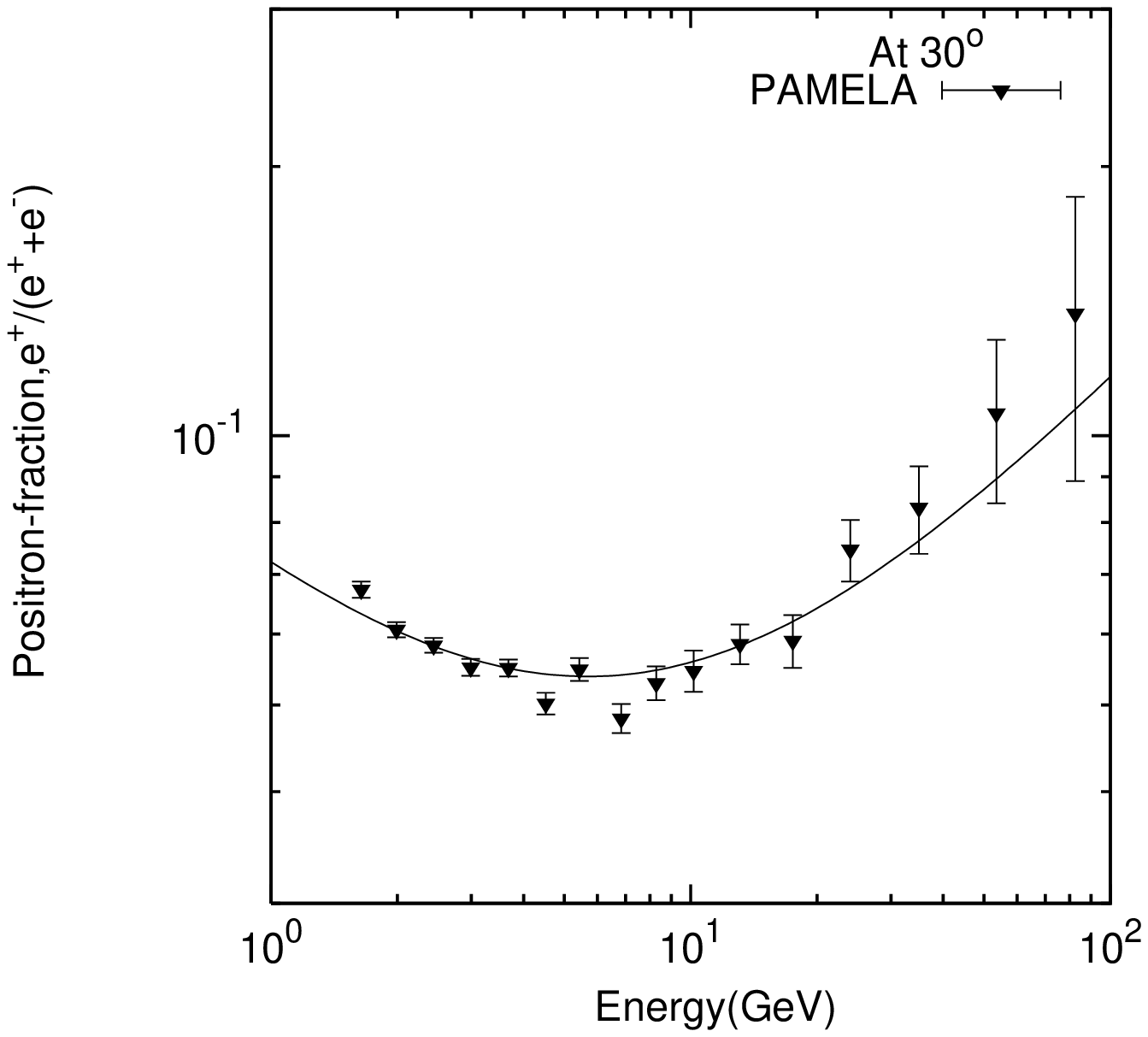}
\setcaptionwidth{2.6in}
\end{minipage}}%
\subfigure[]{
\begin{minipage}{0.5\textwidth}
\centering
 \includegraphics[width=2.5in]{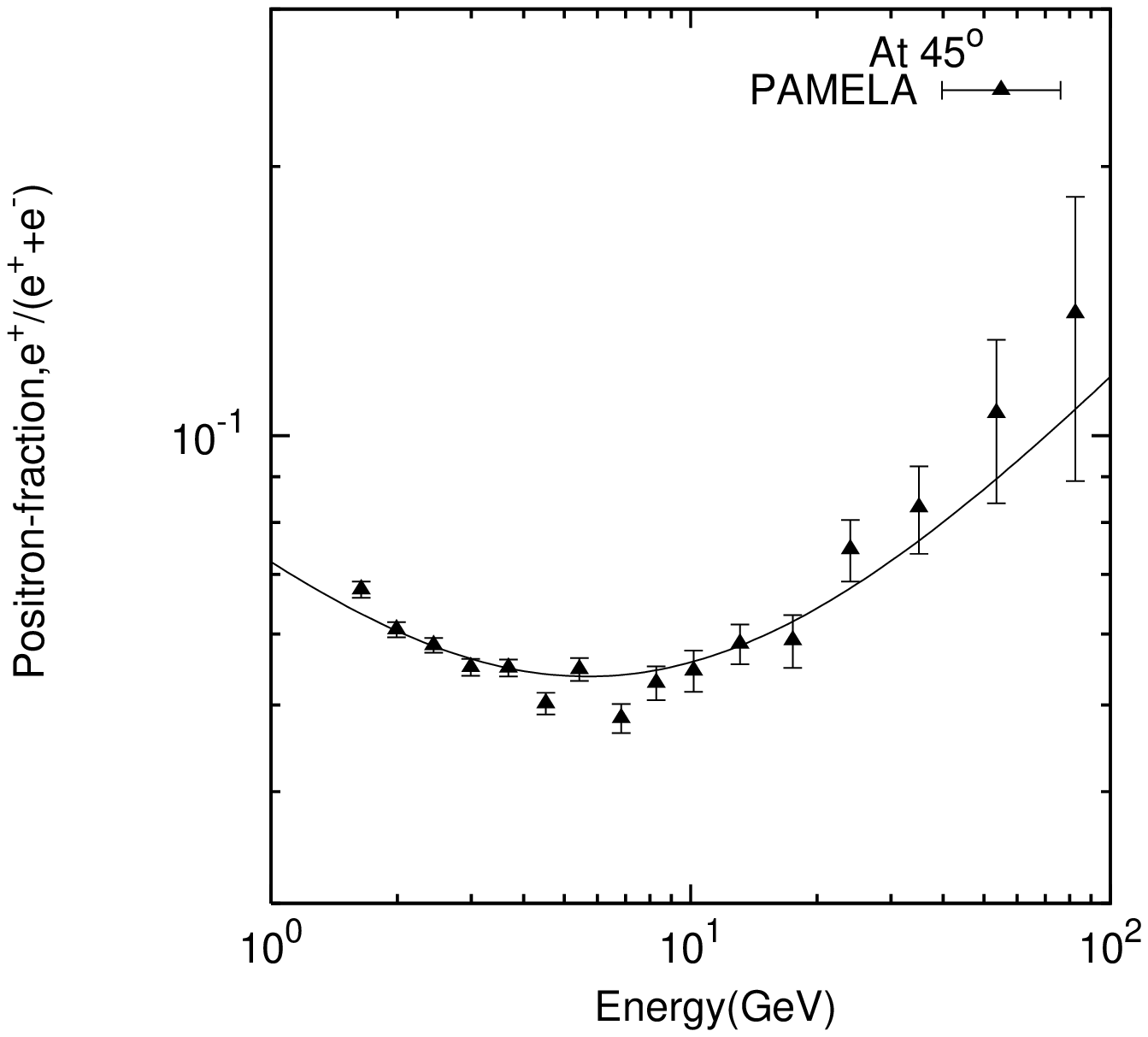}
 \end{minipage}}%
\vspace{0.01in} \subfigure[]{
\begin{minipage}{0.5\textwidth}
\centering
\includegraphics[width=2.5in]{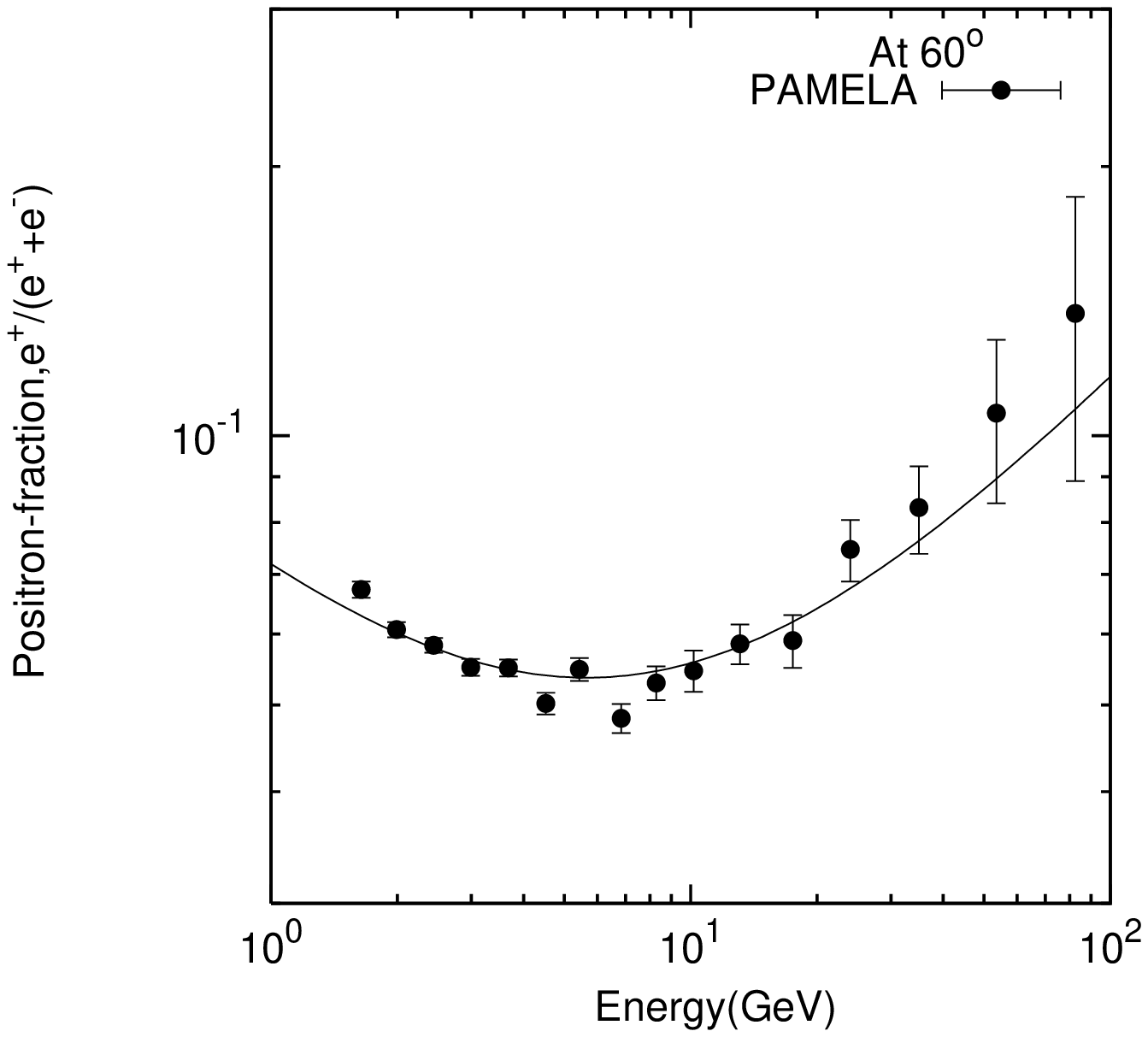}
\end{minipage}}%
\subfigure[]{
\begin{minipage}{.5\textwidth}
\centering
 \includegraphics[width=2.5in]{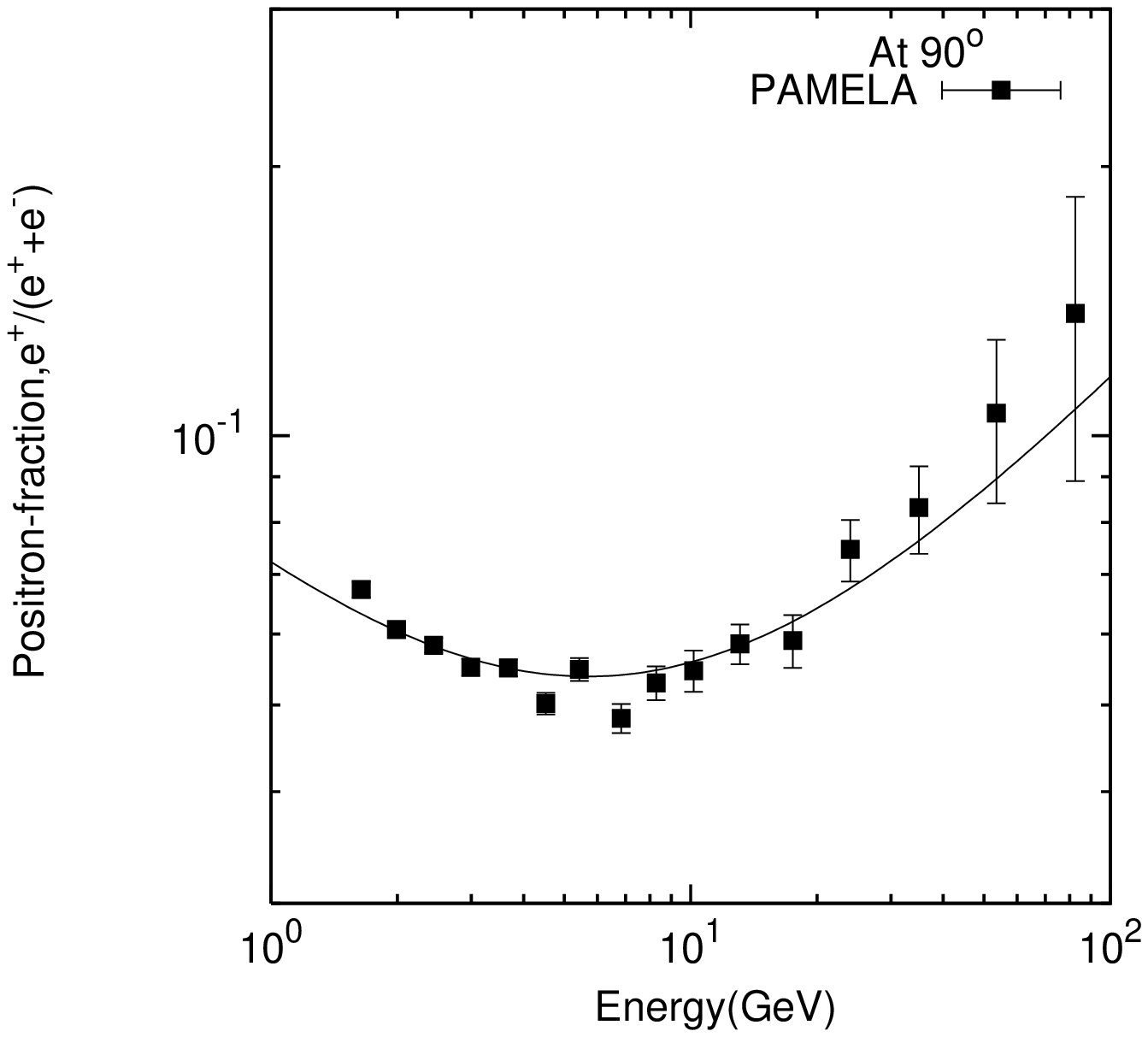}
 \end{minipage}}%
\caption{Data points are taken from Ref.\cite{Adriani1}. The errors are only statistical. The solid curve shows
the results based on the present working formula $(eqn.1)$.}
\end{figure}
\begin{table}[t]
\begin{center}
\begin{small}
\caption{Chosen numerical values of the fit parameters in the
expression for the positron fraction[with large-angle emission].}
\begin{tabular}{|c|c|c|c|}\hline
 $\theta_{cut}$ & $c'$ & $\beta$ & $\frac{\chi^2}{ndf}$\\
 \hline
 $30^{o}$ & $57.248\pm 0.0556$ &$0.011\pm0.0002$  & $3.271/10$\\
 \hline
  $45^{o}$ & $28.632 \pm 0.278$ & $0.011\pm0.0002$  & $3.271/10$ \\
 \hline
  $60^{o}$ & $19.223 \pm 0.183$ & $0.011\pm0.0002$  & $4.242/11$ \\
 \hline
  $ 90^{o}$ & $14.312 \pm 0.139$ & $0.011\pm0.0002$  & $3.271/10$ \\
 \hline
\end{tabular}
\end{small}
\end{center}
\end{table}

\section{Summary and Conclusions :}
The excellent agreement between the measurements and our model-based
results are quite evident in all the cases of assumed both
small-angle or large-angle emissions of the positrons in the PAMELA
experiment. So the anomaly is resolved by virtue of the calculations
done on the basis of the conjectures made here which mark a radical
departure from the conventional set-patterns of ideas and the fixed
standard beliefs. By the same token of argumentative points entailed
in the new paradigm we also predict here the detection of the excess
of cosmic muon neutrinos and antineutrinos at very very high
energies due to the breaking of the nucleons by the cosmic
spallation process in the integergalactic space.
\par The muon charge ratios at very high cosmic ray energies depict
normally a very slowly rising nature. Had the muons been not very
decay-prone, it would also have met the similar nature arising out
of the same nucleon-breaking mechanism. In any case, both the
$\mu^+/\mu^-$ and $e^+/e^-$ ratios rise with energy at very high
cosmic ray energies. This illustrates and manifests one important
aspect of the muon-electron universality property : Our success in
interpreting the PAMELA-data reinforces our dependence on models
different form the Standard Model(s) in the Particle Physics and the
related fields, as the entire edifice of the present work is based
on the rejection of the Standard Model which is artificial and
thoroughly questionable from the very start. However, some other
comments are in order here for the sake of completeness and
totality. (i) The electron-positrons are charged particles. We have,
however, neglected here the complications of path-deflections
suffered by the charged positrons or electrons arising out of the
earth's magnetic field. It is to be stressed upon that the mechanism
suggested here plugs automatically the parallel expectations for the
cases of antiprotons, as the nucleon-(or, proton-)breaking mechanism
can and does in no way give rise to any excess production of
antiprotons. (ii) As it is a ratio, the systematic uncertainties in
the data are cancelled; that also allows the cancellation of the
solar activity. In reality solar modulations could arise from the
phase of the solar cycle and also from the opposite charges of
electrons and positrons. But the fact is we do not include these
probable disturbing effects in order to escape initially the
complications contributing to some very minor effects with no
significant numerical values.
\par However, the model we apply here
takes care of the experimental data quite well and successfully
solves the anomaly evinced by the adjoining figures, without
assuming, however, the role of pulsars, the dark matter
annihilations or taking into account some other phenomena evolved
from the Standard Model(SM)-based points of view. And the conclusion
derived from this work is in perfect accord with what is maintained
by Morselli and Moskalenko\cite{Morselli1} that the excesses imply a
source, conventional or exotic, of additional leptonic component,
especially of the positive variety. And this obviously causes a
distantiation from the viewpoints expressed by Chen et
al\cite{Chen1}.
\par {\bf NOTE ADDED : }
\par After the completion of our present work on PAMELA-Paradox, our
attention was drawn by a scientific colleague to a very important
report\cite{Butt1} and to a few papers\cite{Chang1,Aharonian1} with
some concrete findings. The report by Butt\cite{Butt1} is an
exceptionally brilliant one for its unbiasedness or open-mindedness
and near-perfect objectivity. The bump observed by the ATIC
collaboration is not to be electron-specific; side by side with the
signature-electrons, the positrons are also to be produced in
roughly equal measure, as was expressly maintained by
Butt\cite{Butt1}, if Kaluza-Klein(KK) WIMPs are their progenitor.
But the ATIC measurements\cite{Chang1} were singularly aimed at
studying the electrons alone. So, unless the positrons are measured
under the same or similar circumstances by these
groups\cite{Chang1,Aharonian1}, no definitive comment on the status
of KK WIMPs is possible. In this context, another comment is yet in
order. The measurements by Aharonian et al\cite{Aharonian1} for
detection of cosmic ray electrons at energies beyond 600GeV do not
report very clearly, such excesses, as is indicated by Chang et al
\cite{Chang1}. This, therefore, tacitly and indirectly puts a
question mark to the measurements of Chang et al until further
scrutiny.
\par
Besides, there are a few sharp differences between the two sets of
studies[PAMELA, HEAT etc. on the one side, and ATIC, HESS etc., on
the other], for which a comparison of the two sets of findings might
not be quite significant. Firstly, the studies by PAMELA
Collaboration concentrate uniquely on the studies of cosmic
antiparticles like positron(s) and antiproton(s), whereas both the
ATIC and the HESS Collaborations measure only electron spectrum at
much higher energies; they did not report on the measure of
corresponding antiparticle production scenario. Secondly, the
energy-ranges of the two sets of experiments are grossly different.
So the studies on the excess production of cosmic electrons alone at
much higher energies do not disturb the pivot of our present work on
PAMELA- paradox. Unless ATIC and HESS Collaborations study and
report pointedly on the nature of positron and antiproton
productions at the same energy-range and under the stringently same
experimental conditions we see no tangible reason to redefine our
attitude and reconsider our outlook. However, if asymmetric excess
electron production alone is repeatedly reported and confirmed, only
then we will have to investigate in to some exotic sources or
nor-yet-proposed-or-known mechanism for cosmic electron production.
So for the present, we are still not in favour of attaching too much
importance to the hypothesis of dark matter or the postulates
Kaluza-Klein(KK) WIMPs.
\par{\bf Acknowledgements :} The authors express their thankful
gratitude to an anonymous Referee for some encouraging remarks and
helpful comments.

\end{document}